\documentclass[aps,prl,twocolumn,showpacs]{revtex4}
\usepackage{graphicx} 
\begin{document}
\title
{\bf Unconventional pairing in exciton condensates under spin-orbit coupling}
\author{M. Ali Can and T. Hakio\u{g}lu}
\affiliation{
Department of Physics, Bilkent University, 06800 Ankara, Turkey}

\begin{abstract}
It is shown that the Rashba and Dresselhaus spin orbit couplings enhance the conclusive power in the
experiments on the excitonic condensed state by at least three low temperature effects. First, spin 
orbit coupling facilitates the photoluminescense measurements via enhancing the bright contribution 
in the otherwise dominantly dark exciton condensed state. The second is the presence of a power law 
temperature dependence of the 
thermodynamic observables in low temperatures and the weakening of the second order transition at 
the critical temperature. The third is the appearance of the nondiagonal elements in the static spin 
susceptibility. 
\end{abstract}
\pacs{71.35.-y,71.70.Ej,03.75.Hh,03.75.Mn}
\maketitle

The existence of low temperature excitonic collective states in
bulk semiconductors has been speculated in the early 1960s.\cite{SAM}
 The idea was that, due to the attractive
interaction between the electron-hole pairs, excitons act like single   
bosons in low densities and consequently,
they should condense in 3D at sufficiently low temperatures. The
difference of the speculated excitonic Bose Einstein condensate (BEC) from the
atomic BEC is that, due to the small exciton band
mass ($m_x\simeq 0.07 m_e$ where $m_e$ is the bare electron mass)
the condensation is expected to occur at much higher critical temperatures
than the atomic BEC. One of the first experimental
results were on bulk $CuO_2$ samples.\cite{exp1,Fukuzawa}
In the last 15 years however experimental efforts were primarily focused on
coupled quantum wells (CQW). The main motivation there is that, spatially indirect 
excitons can have much longer lifetimes (in the order of $10 \mu s$) in the 
presence of an additional electric field applied perpendicular to the plane, in 
contrast with the bulk where lifetimes are on the order of $ns$. 
Recent reviews\cite{Butov_2007} with CQWs remark the 
evidence of a low temperature exciton condensate (EC) in the observations
of large indirect exciton mobility and radiative decay rates, enhancements in 
exciton scattering rate, and the narrowing of the photoluminescense (PL) spectra. 
It is also suggested that there is room for different conclusions other than the excitonic 
BEC\cite{Butov_2007,Snoke_rev}.

Here we propose an alternative method to examine the EC by breaking the
 spin degeneracy of the electrons (e) and holes (h) via a spin-orbit coupling (SOC)
of the Rashba\cite{RSOC} (RSOC) and the Dresselhaus\cite{DSOC} (DSOC) types. The RSOC is
manipulated by the external electric (E)-field whereas the DSOC is known to be intrinsically
present in zinc-blende structures\cite{Winkl_book}. There are major differerences
between the SOC induced effects
in the non-centrosymmetric superconductors and the semiconductor electron hole (e-h) CQWs. In
the former the SOC is strong relative to the typical condensation energy, i.e.
$\Delta_0 \lesssim E_{so} \ll E_F$ where $E_{so}$ and $\Delta_0$ are typical SOC and condensation
energies\cite{GR_2001} and $E_F$ is the Fermi energy;
whereas in the latter $E_{so} < \Delta_0 \ll E_F$\cite{HS}. It may thus seem that such a perturbative
effect may not play a significant role in the thermodynamics of the latter case. However, in an
otherwise spin degenerate EC, the SOC changes the nature of the ground state as well
as the low temperature behaviour of
the thermodynamic observables. Three distinct effects of the SOC are expected in the EC as,
i) a controllable mixture of the dark and bright condensates (DC and BC hereon) in the
ground state, ii) finite off diagonal static spin susceptibilities, and iii)
non isotropic narrowing between the two lowest energy bands near $E_F$ manifesting in the low
temperature behaviour of the thermodynamic variables.

The reasons for the lack of conclusive evidence for the EC in the PL measurements was recently
suggested in relation to (i) above.\cite{Combescot2} It was proposed that the Pauli exclusion
principle in the presence of interband Coulomb scattering implements a ground state predominantly composed
of DC which does not couple to light due to its spin $(\pm 2)$. It is claimed that
it is mainly the BC [spin $(\pm 1)$] that is probed in the PL experiments which is
energetically above the DC by approximately $0.1 \mu eV$. Here, we take the relevant 
fundamental symmetries or some of their breakings, [i.e. spin degeneracy (S), reflection symmetry (P), 
rotational invariance (R) of the Coulomb interaction, time reversal symmetry (T) and the fermion exchange 
symmetry (FX)], into full account. It is confirmed that the ground state is dominated 
by the dark condensate in the absence of the SOC. In most semiconductors, the SOC is in the range
$0.1-10 meV$ for RSOC and $1-30 meV$ for the DSOC with the larger ones being for the indium based materials at
typical exciton densities $n_x \lesssim 10^{11} cm^{-2}$. We demonstrate that a weak SOC can change the
groundstate from dominantly DC to a controllable mixture of DC and BC\cite{wang}. 
This is crucial in the PL experiments that such observations are expected to be more conclusive 
than detecting the DC indirectly by its influence on the line shift of the BC.\cite{Combescot2} 
Additionally, and in relation
to (ii), the SOC permits the off diagonal spin susceptibility to be finite, enabling
complementary measurements to PL. 
In relation to (iii), there are additional finite temperature effects of the SOC on the band structure 
reminiscent of the zero temperature lines-of-nodes observed in the superconducting gap of the non-centrosymmetric superconductors. 
There, due to the anisotropy of the strong SOC, the zero temperature superconducting
gap vanishes at certain lines on the Fermi surface changing the temperature behaviour of the thermodynamic 
quantities from exponentially suppressed to power law.\cite{Samokhin_05} In the present case, a similar effect 
at low temperatures is produced in the presence of the SOC.   

We demonstrate these three effects (i-iii) by considering a model of e-h type CQWs confined
in the $x-y$ plane with $z$ [ $(0,0,1)$ of the underlying
lattice] as the growth direction. The CQWs are separated by a tunneling barrier of thickness $d \simeq 100 \AA$
and the width of each well is $W= 70 \AA$.
The model Hamiltonian is
\begin{eqnarray}
{\cal H}_X&=&\sum_{{\vec k},\sigma,p} (\zeta_{\vec k}^{(p)}-\mu_p)
\hat{p}^\dagger_{{\vec k},\sigma}\hat{p}_{{\vec k},\sigma}+{\cal H}_{so}^{(e)}+{\cal H}_{so}^{(h)} \nonumber \\
&+&\frac{1}{2}\sum_{{{\vec k},{\vec q} \atop \sigma,\sigma^\prime,p,p^\prime}}
V_{pp^\prime}(q)\hat{p}^\dagger_{{\vec k}+{\vec q},\sigma}\hat{p^\prime}^\dagger_{{\vec k}^\prime-{\vec q},\sigma^\prime}
\hat{p^\prime}_{{\vec k}^\prime,\sigma^\prime}\hat{p}_{{\vec k},\sigma}
\label{full_hamilt}
\end{eqnarray}
where $\mu_p$ is the chemical potential for electrons ($p=e$) and heavy holes ($p=h$). Here $\hat{p}_{{\vec k}\uparrow}$
indicates the annihilation operator at the wavevector ${\vec k}$ and spin=+1/2 for
$p=e$, and spin=+3/2 for $p=h$. The spin independent {\it bare} single particle energies are 
$\zeta_{\vec k}^{(p)}=\hbar k^2/(2m_p^*)$
where $m_p^*$ is the effective mass ($m_e^*=0.067 m_e$ and $m_h^*=0.4 m_e$ where $m_e$ is the free electron mass) and 
$V_{pp^\prime}(q)=s_{pp^\prime}2\pi e^2/(\epsilon_{\infty} q)$ is the Coulomb
interaction between $p$ and $p^\prime$ particles where $s_{ee}=s_{hh}=1$ and $s_{eh}=-2$. The CQWs are formed 
within the GaAs-AlGaAs interface. The electron
R$\&$D SOCs are described by ${\cal H}_{so}^{(e)}$ as
[in the basis $(\hat{e}_{{\vec k}\uparrow}~ \hat{e}_{{\vec k}\downarrow})$],
\begin{equation}
{\cal H}_{so}^{(e)}({\vec k})=\pmatrix{0& i\alpha_R^{(e)} k_- +\alpha_D^{(e)} k_+ \cr
-i\alpha_R^{(e)} k_+ +\alpha_D^{(e)} k_- & 0}
\label{Rashba_full}
\end{equation}
where $\alpha_R^{(e)}=r^{6c6c}_{41}E_0$ and $\alpha_D^{(e)}=-b^{6c6c}_{41} \langle k_z^2\rangle$ with ${\vec E}=E_0 {\vec e}_z$
as the external E-field, and $r^{6c6c}_{41}\simeq 5.2 e\AA^2, b^{6c6c}_{41}\simeq 27.6 eV\AA^3$ are the
coupling constants for GaAs\cite{Winkl_book} with $\langle k_z^2 \rangle\simeq (\pi/W)^2$.
The hole R$\&$D SOCs are described by ${\cal H}_{so}^{(h)}$ as
[in the basis $(\hat{h}_{{\vec k}\uparrow}~ \hat{h}_{{\vec k}\downarrow})$],
\begin{equation}
{\cal H}_{so}^{(h)}({\vec k})=\pmatrix{0 & \alpha_R^{(h)} k_-^3 + \alpha_D^{(h)} k_+ \cr
\alpha_R^{(h)} k_+^3 +\alpha_D^{(h)} k_- & 0}
\label{Dresselhaus_full}
\end{equation}
where $\alpha_R^{(h)}=\beta_h E_0$ and $\alpha_D^{(h)}=-b^{7v7v}_{41}\langle k_z^2\rangle$ with
$\beta_h \simeq 7.5 \times 10^{6} e\AA^4, b^{7v7v}_{41} \simeq -58.7 eV\AA^3$ as the
coupling constants for the holes\cite{Winkl_book}.
In the combined e-h basis $( \hat{e}_{{\vec k}\uparrow}~ \hat{e}_{{\vec k}\downarrow} ~
\hat{h}_{-{\vec k} \uparrow}^\dagger ~\hat{h}_{-{\vec k} \downarrow}^{\dagger})$, the
condensed system has the reduced $4\times 4$ Hamiltonian (up to a multiple of the unit matrix)
\begin{equation}
{\cal H}_X=\pmatrix{
{\bf \Sigma}^{(e)}({\vec k})-\mu_x & {\bf \Delta}^\dagger({\vec k})\cr
{\bf \Delta}({\vec k}) & {\bf \Sigma}^{(h)}(-{\vec k})+\mu_x}
\label{hamilt_2by2}
\end{equation}
where $\mu_x$ is the exciton chemical potential found self consistenly by conserving the number of exciton pairs\cite{HS,Zhu} and
the diagonals are the e and h single particle energies 
\begin{equation}
{\bf \Sigma}^{{e \choose h}}({\vec k})=\pmatrix{\pm \tilde{\zeta}_{{\vec k} \uparrow} & \pm \Sigma_{\uparrow \downarrow}^{{e \choose h}*}({\vec k})+{\cal H}_{so}^{{e \choose h}}({\vec k})\cr
\pm \Sigma_{\uparrow \downarrow}^{{e \choose h}}({\vec k})+{\cal H}_{so}^{{e \choose h}*}({\vec k}) & \pm \tilde{\zeta}_{{\vec k} \downarrow} }~.
\label{sigma_e}
\end{equation}
\begin{table}
\begin{tabular}{l|l}
\hline
Symmetry & Action  \\
\hline T & ${\bf X}({\vec k}) \to (-i \sigma_y) {\bf X}^*(-{\vec k}) (-i \sigma_y)^{-1}$   \\
 S  & $\sigma_x {\bf \Delta}({\vec k})\sigma_x$    \\
 P  & ${\bf X}({\vec k}) \to {\bf X}(-{\vec k})$    \\
 FX  & ${\bf \Delta}({\vec k}) \to -{\bf \Delta}^T(-{\vec k})~,
{\bf \Sigma}^{(e)}({\vec k}) \leftrightarrow {\bf \Sigma}^{(h)}({\vec k})$ \\
\hline
\end{tabular}
\caption{The fundamental symmetry operations. Note that FX is inapplicable in this work. Here 
$\sigma_x, \sigma_y$ are the Pauli-Dirac matrices and 
${\bf X}={\bf \Sigma}^{(p)},{\bf \Delta}$ as given by Eq's (\ref{sigma_e}), (\ref{delta_eh}).}
\vspace{-0.5cm} \label{table1}
\end{table}
where $\tilde{\zeta}_{{\vec k} \sigma}=[\tilde{\zeta}_{{\vec k} \sigma}^{(e)}+\tilde{\zeta}_{{\vec k} \sigma}^{(h)}]/2$
includes the diagonal self energies, i.e.  
$\tilde{\zeta}_{{\vec k} \sigma}^{(p)}=\zeta_{\vec k}^{(p)}+\Sigma_{\sigma \sigma}^{(p)}({\vec k})$. The off diagonal components
$\Sigma_{\uparrow \downarrow}^{(p)}=\Sigma_{\downarrow \uparrow}^{(p) *}$ in Eq.(\ref{sigma_e})
 represent the self energies of cross spin correlations.
The non diagonal elements in Eq(\ref{hamilt_2by2}) are the spin dependent excitonic order parameters
\begin{equation}
{\bf \Delta}({\vec k})=\pmatrix{
\Delta_{\uparrow \uparrow}({\vec k}) & \Delta_{\downarrow \uparrow}({\vec k})\cr
\Delta_{\uparrow \downarrow}({\vec k}) & \Delta_{\downarrow \downarrow}({\vec k})}
\label{delta_eh}
\end{equation}
with the diagonals corresponding to the DC\cite{Combescot2} and the off diagonals
corresponding to the BC in the mixture of {\it triplet}\cite{triplet_singlet} (T) and the {\it singlet} (S) state
$\Delta_{T \choose S}({\vec k})=[\Delta_{\uparrow \downarrow}({\vec k})
\pm \Delta_{\downarrow \uparrow}({\vec k})]/2$.

The solution of Eq.(\ref{hamilt_2by2}) includes the self consistent mean field Hartree-Fock calculation of the
thermodynamic state where 
\begin{eqnarray}
\Sigma^{(p)}_{\sigma \sigma^\prime}({\vec k})&=&-\sum_{\vec q} V_{pp}({\vec q})
\langle \hat{p}_{{\vec k}+{\vec q} \sigma}^\dagger \hat{p}_{{\vec k}+{\vec q} \sigma^\prime}\rangle~,\quad \hat{p}=(\hat{e},\hat{h}) \nonumber \\
\Delta_{\sigma \sigma^\prime}({\vec k})&=&\sum_{\vec q} V_{eh}({\vec q}) \langle \hat{e}_{{\vec k}+{\vec q},\sigma}^\dagger
\hat{h}_{-{\vec k}-{\vec q},\sigma^\prime}^\dagger \rangle
\label{mean_field}
\end{eqnarray}

\begin{figure}[t]
\includegraphics[scale=0.33,angle=0]{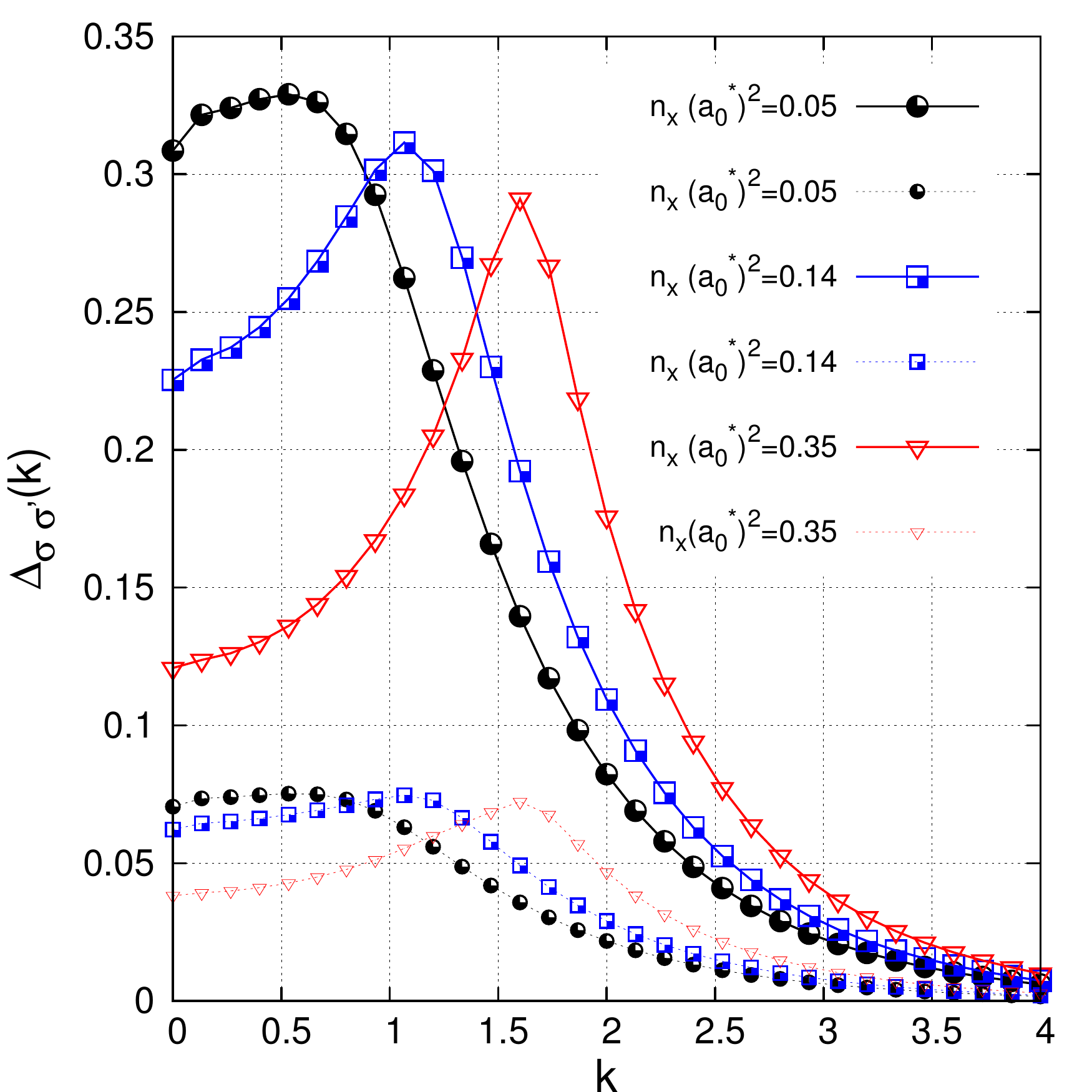}
\caption{(Color online) The non vanishing zero temperature order
parameters for DC (i.e. $\Delta_{\uparrow
\uparrow}=\Delta_{\downarrow \downarrow}$), and BC (i.e.
$\Delta_{\uparrow \downarrow}$) in the spin independent case
corresponding to three densities ranging from BEC [i.e. $n_x
(a_0^*)^2 < 0.1$] to BCS [i.e. $0.1 < n_x (a_0^*)^2$] type. Solid
and dashed lines respectively correspond to DC and BC at the
corresponding exciton density.} \vspace{-0.5cm}
\label{dark_bright_no_soc}
\end{figure}
with $\langle \dots \rangle$ indicating the thermal average.
The fundamental symmetry operations affecting the solution of Eq(\ref{hamilt_2by2}) are R, T, S and P. 
As a fundamental difference of the EC in CQWs from the non-centrosymmetric superconductors, the 
fermion exchange (FX) symmetry is absent in the former manifesting 
in the appearance of the "triplet" states even in the presence of T symmetry and, the order parameters 
with a fixed total spin have mixed parities. 
The transformations corresponding to these symmetries are shown in Table.I. 
In the absence of SOC the respected symmetries are R,T,S,P, whereas with finite SOC only T is manifest.
An elementary application of Table.I in the absence of SOC and FX shows that  
$\Sigma_{\uparrow \downarrow}^{(p)}({\vec k})=\Sigma_{\downarrow \uparrow}^{(p)^*}({\vec k})=0$, and  
  $\Sigma_{\uparrow \uparrow}^{(p)}({\vec k})=\Sigma_{\downarrow \downarrow}^{(p)}({\vec k}) \ne 0$.
It is also required that,
$\Delta_{\uparrow \uparrow}({\vec k})=\Delta_{\downarrow \downarrow}({\vec k}) \ne 0$ for the DC whereas
the BC in the triplet state vanishes, i.e. $\Delta_T({\vec k})=0$ therefore $\Delta_S({\vec k})=\Delta_{\uparrow \downarrow}({\vec k})$. 
Due to the rotational symmetry of the Coulomb interaction all finite elements are real.    
The radial configurations of the isotropic DC and BC order parameters are depicted in Fig.(\ref{dark_bright_no_soc}) 
indicating that the the bright "singlet" is much weaker than the dark "triplets" and the ground state is dominated by the DC.

\begin{figure}[t]
\includegraphics[scale=0.4,angle=0]{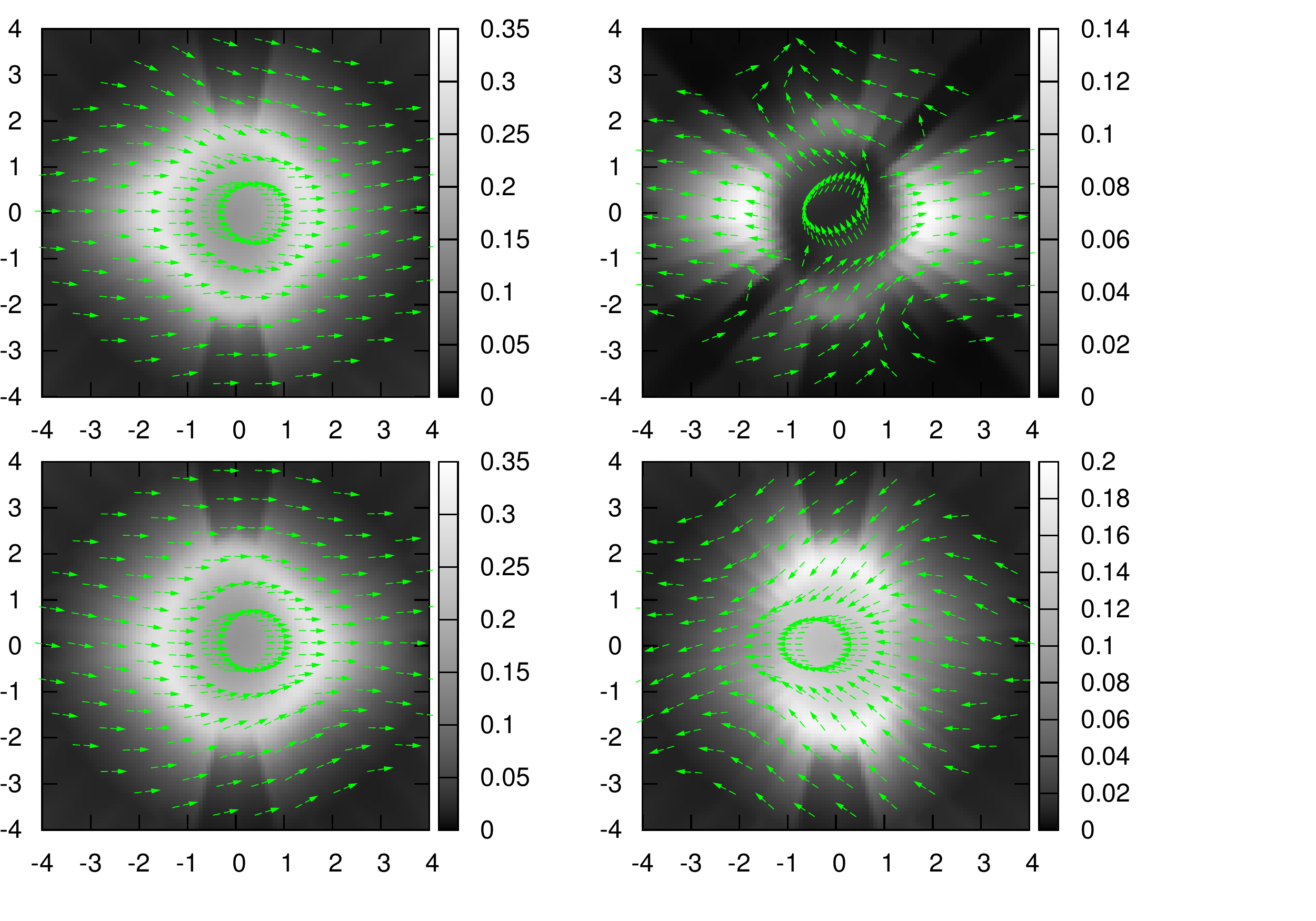}
\vspace{-0.5cm} \caption{(Color online) The DC and BC as function
of ${\vec k}$ in the presence of R$\&$D SOC with the same
parameters used in Fig.(\ref{dark_bright_no_soc}). Starting from
the top left in counterclockwise, $\Delta_{\uparrow \uparrow}$,
$\Delta_{\downarrow \downarrow}$, $\Delta_{S}$, $\Delta_{T}$. Gray
scale is the magnitude and arrows depict the phase. Here $E_0=75
kV/cm$, $n_x (a_0^*)^2=0.35$ and the SOC parameters used are as
given below Eq.(\ref{Rashba_full}) and
Eq.(\ref{Dresselhaus_full}).} \vspace{-0.5cm}
\label{dark_bright_w_soc}
\end{figure}

A more interesting case is when the R$\&$D SOCs are present. With the broken R,S,P,FX and manifest T symmetries   
Eq.(\ref{delta_eh}) is a complex matrix with unconventional phase texture particularly in the cross  
spin configurations. 
The FX breaking is even stronger due to the different couplings in the conduction and the valence bands. 
 Fig.(\ref{dark_bright_w_soc}) depicts the dark and bright components of
Eq.(\ref{delta_eh}) in the presence of R$\&$D SOCs for GaAs with $E_0=75 kV/cm$. It is observed that 
$\Delta_T$ and $\Delta_S$ have comparable magnitudes to those of the DC.

\begin{figure}[t]
\includegraphics[scale=0.39,angle=-90]{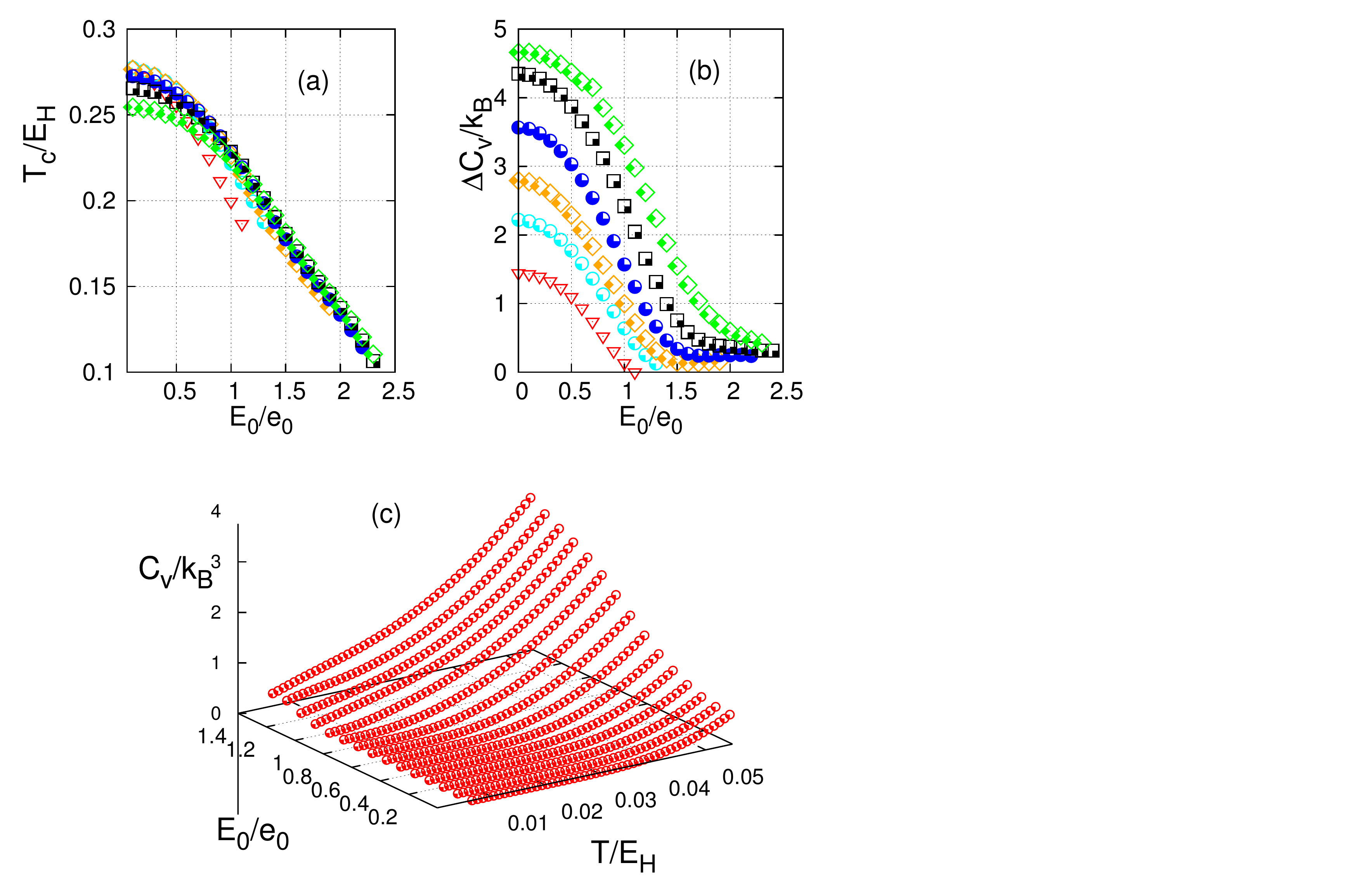}
\caption{(Color online) (a) $T_c/E_H$ vs $E_z$, (b) $\Delta C_v$ vs. $E_0$, 
and (c) $C_v$ vs. $T$ and $E_0$ at $T \ll T_c$ for $n_x {(a_0^*)}^2=0.29$. Here  
$e_0=150 kV/cm$, $E_H$ is the Hartree energy and $k_B$ the Boltzmann constant. 
Symbol coding at different $n_x$ is shared between (a) and (b). From the top to bottom plots in (b) 
the densities are $n_x {(a_0^*)}^2=0.29,0.33,0.38,0.42,0.46,0.62$.} 
\vspace{-0.5cm} \label{cv_sus_1}
\end{figure}

We have thus far shown that the SOC enhances the BC in the ground state in the context of the first effect (i) above.
We now examine the second and the third effects (ii) and (iii) together in the temperature and SOC dependence of the specific heat $(C_v)$ 
and the non diagonal spin susceptibility $(\chi_{xz})$. In the absence of SOC, the temperature dependence of $C_v$ is exponentially 
suppressed for $T\simeq 0$. The critical temperature $T_c$ is identified by the $\lambda$-anomaly $\Delta C_v$ where the 
condensate has a second order transition into the e-h liquid.   
 Increasing the SOC decreases $T_c$ [Fig.(\ref{cv_sus_1}.a)], weakens the anomaly at the $T_c$ [Fig.(\ref{cv_sus_1}.b)]  
and changes the behaviour to a power law near $T \simeq 0$ [as shown for $n_x {(a_0^*)}^2=0.29$ in Fig.(\ref{cv_sus_1}.c)]. 
At low densities where the condensation is stronger, the exponential suppression is more robust 
to changes into a power law than in the case of high densities. The critical temperature $T_c$ is weakly density  
dependent whereas $\Delta C_v$ is much stronger at low densities [Fig.(\ref{cv_sus_1}).a and b]. 
The overall effect is that $\Delta C_v/T_c$ is not a  
universal ratio, i.e. larger in the strongly interacting low density BEC limit than in the weakly interacting 
high density BCS limit. This ratio is reduced as the SOC is increased (more rapidly at high densities) implying that the 
strongly second order transition is weakened by increasing the SOC. In the high density range, ${\Delta C}_v$ vanishes at 
a certain SOC strength whereas at low densities the second order transition remains but dramatically weakened. It is expected 
that the Hartree-Fock scheme studied here may be insufficient in explaining the details of the transition due to 
the strong dipolar fluctuations at high densities and near $T_c$. 
Two factors are influencial on the $C_v$ near $T\simeq 0$ and $T\simeq T_c$ when the SOC strenght is increased. 
The first is the anisotropic narrowing of the two lowest energy exciton bands at the Fermi energy allowing low energy 
thermal excitations between these bands. Thermal interband excitations are affective even at low temperatures
resulting in a power law dependence on $T$. The second is that the anisotropy of the complex order parameters 
persists at finite temperatures which is more affective in larger   
densities averaging out the anomalous contribution to $C_v$ more rapidly than in low densities. 
\begin{figure}[b]
\vspace{-1cm}
\includegraphics[scale=0.32,angle=0]{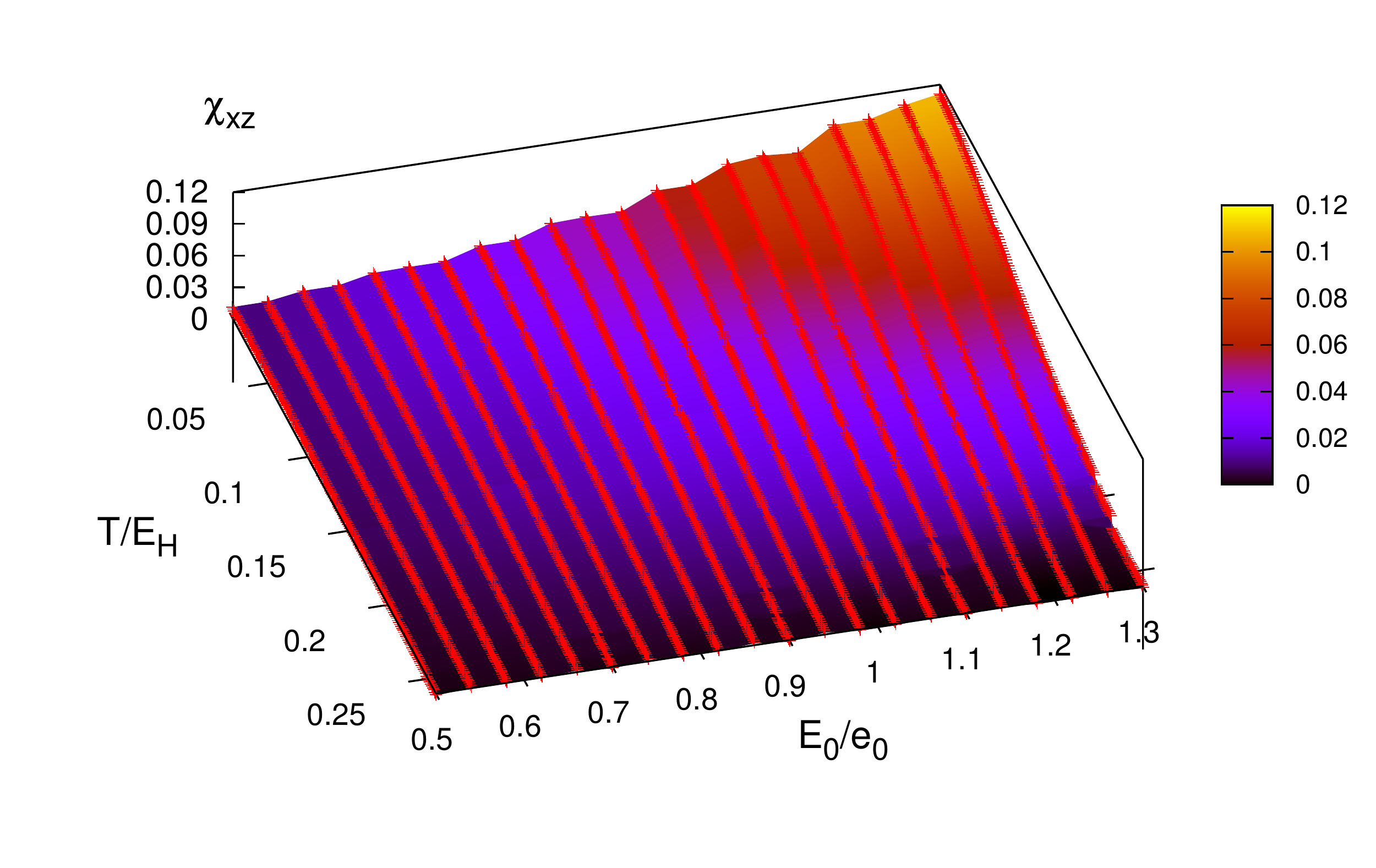}
\caption{(Color online) The non diagonal component $\chi_{xz}$ 
 as a function of $T$ and $E_0$ at $n_x{(a_0^*)}^2=0.29$.} 
\label{nd_sus_1}
\end{figure}

Concerning $\chi_{xz}(T)$, it vanishes at all temperatures in the absence of SOC. However, it is finite in the presence 
of SOC\cite{HS} and stronger in low temperatures as depicted in Fig.(\ref{nd_sus_1}). For increasing SOC,    
$\chi_{xz}(0)$ gradually increases and an exponential-to-power law behaviour near $T \simeq 0$ develops 
similar to $C_v$ in Fig.(\ref{cv_sus_1}.c). 

In conclusion, we demonstrated three measurable effects displayed by an e-h system when the R$\&$D  
SOCs are considered. The first is the enhancement of the BC in the ground state
with observable consequences in the PL measurements. The second and the third are the non conventional 
behaviour of the specific heat and the spin susceptibility at $T\simeq 0$ and $T=T_c$. These three 
effects combined should help in the enhancement of our understanding the exciton condensation. 

Recent experiments\cite{stress} have shown that the broken inversion symmetry in semiconductor  
heterostructures can yield, in addition to the R$\*$D SOCs, strain induced spin splitting. 
The splitting observed in these systems are in the same order of magnitude as the SOC studied here\cite{BZ}.  
We expect the PL and the thermodynamic measurements under strain to be crucially important for producing  
complementary results to the PL measurements.  

This research is funded by T\"{U}B{\.I}TAK-105T110. We thank
Marmaris Institute of Theoretical and Applied Physics (ITAP) where parts of this work were completed.


\begin{thebibliography}{99}
\bibitem{triplet_singlet} Terminology here is coined by the analogy with the identical spin-1/2 
pairings. In our case, these components correspond to the total spin (S=2,1) due to 
the electron ($S_e=1/2$) and the heavy hole orbitals ($S_h=3/2$).
\end{thebibliography}
\end{document}